\documentclass[12pt,preprint]{aastex}

\newcommand{\be}{\begin{equation}}
\newcommand{\ee}{\end{equation}}

\shorttitle{Neutral gas density in DLA systems}
\shortauthors{Trenti \& Stiavelli}

\begin{document}


\title{Neutral gas density in Damped Lyman $\alpha$ systems}


\author{M. Trenti \and M. Stiavelli}
\affil{Space Telescope Science Institute, 3700 San Martin Drive Baltimore MD 21218 USA}
\email{trenti@stsci.edu; mstiavel@stsci.edu}


\begin{abstract}
\emph{Accepted for publication in the Astrophysical Journal\\}
We estimate the intrinsic neutral gas density in Damped Lyman $\alpha$
systems ($\Omega_{HI}^{(DLA)}$) in the redshift range $ 2.2 \lesssim z
\lesssim 5$ from the DLA SDSS DR\_3 sample of optically selected
quasars.  We take into account self-consistently the obscuration on
background quasars due to the dust present in Damped Lyman $\alpha$
systems. We model the column density and redshift distribution of
these systems by using both a non-parametric and a parametric
approach.  Under conservative assumptions on the dust content of
Damped Lyman $\alpha$ systems, we show that selection effects lead to
underestimating the \emph{intrinsic} neutral gas density by at least
$15\%$ with respect to the \emph{observed} neutral gas density. Over
the redshift range $[2.2;5.5]$ we find
$\Omega_{HI}^{(DLA)}=0.97^{+0.08+0.28}_{-0.06-0.15} \cdot
10^{-3}$, where the first set of error bars gives the $1\sigma$ random
errors and the second set gives the modeling uncertainty dependent on
the fraction of metals in dust - from 0\% to 50\%. This value compares
with $\Omega_{HI}^{(DLA)}=0.82^{+0.05}_{-0.05}$ ($1\sigma$ error bars),
which is obtained when no correction for dust is introduced. In the
model with half of the metals mass in dust we cannot constraint
$\Omega_{HI}^{(DLA)}$ at a confidence level higher than $90\%$. In
this case there is indeed a probability of about $10\%$ that the
intrinsic column density distribution of DLA systems is a power law
$f(N_{HI}) \propto 1/N_{HI}^{~1.95}$. In contrast, with $25 \%$ of the
metals in dust - the most realistic estimate - a power law is ruled
out at $99.5\%$ of confidence level.

\end{abstract}


\section{Introduction}

Damped Lyman $\alpha$ systems (hereafter DLA systems) are quasar
absorption systems with a column density above $2\cdot 10^{20}
cm^{-2}$ and represent the high end of the distribution of absorption
systems starting from the Lyman $\alpha$ forest at $N_{HI} \gtrsim
10^{14} cm^{-2}$. DLA systems represent the most significant reservoir
of neutral hydrogen in the universe available for star
formation. These systems are considered to be either (cold) massive
rotating disks, the progenitors of todays disk galaxies
\citep{pro97,wol05} or compact protogalactic clumps
\citep{hae98,nag04}.

In the era of precision cosmology, an accurate measure of the total
mass density of neutral gas as a function of the redshift represents
an important constraint for galaxy formation models. Ground
observations are able to identify DLA absorption features in the
spectra of quasars from $z_{abs} \gtrsim 1.8$, where the absorption
lines enter the atmospheric window, up to $z_{abs} \approx 5.5$. With
an all sky survey like the Sloan Digital Sky Survey, spectra of
several thousands of quasars with enough resolution for DLA detection
have been acquired and the \emph{observed} density of neutral gas in
DLA systems is now measured with errors below $10\%$ \citep{pro05}.

This measurement must be interpreted with some caution, as the
presence of DLA systems along a line of sight leads to a potential
obscuration due to the dust that they host: the \emph{observed} gas
density is a biased estimator of the \emph{intrinsic} density unless
the dust effects are accurately quantified. Several papers, starting
from \citet{ost84} have attempted to model the influence of dust along
the line of sight, often with conflicting results.

A detailed analysis framework for the obscuration of quasars has been
developed by \citet{fal93} (see also \citealt{fal89,fal89b,pei91,pei95b}) and
applied to the quasars sample of \citet{lan91}. Their study
highlighted a potentially severe effect of the dust bias that did not
allow to put an upper limit to the intrinsic density of DLA
systems. In fact, absorbers with high column densities and/or with
high dust-to-gas ratio represent essentially ``bricks'' along the line
of sight to a quasar and are very likely to be missed in optically
selected surveys. An additional evidence for the dust obscuration came
in the form of a detected preferential reddening in the spectra of
quasars with DLA absorption with respect to a control sample without
detection of these systems \citep{pei91}.

More recent investigations revised these earlier results on the dust
content in DLA systems and on their reddening of background objects
\citep{mur04,ell05}, finding in particular no robust evidence for the
reddening of quasars at $z \approx 3$ with DLA features in their
spectra: at $3\sigma$ \citet{mur04} find $E(B-V) < 0.02mag$, while
\citet{ell05} have $E(B-V)<0.04mag$ . At the same time radio selected
quasars surveys \citep{ell01}, with complete optical follow-up
detection have provided the first bias free constraints on the
intrinsic distribution of DLA systems.

Taking advantage of these recent measurements for the number density
of DLA systems, we have previously characterized \citep{tre06} the
dust absorption along random lines of sight by means of a Monte
Carlo code, finding that, on average, the deviations from unit
transmission are effectively modest ($\langle \exp{(-\tau)} \rangle
\gtrsim 0.9$ at an emitted wavelength $\lambda_e = 0.14 \mu m$ over
all the redshift range) and of limited impact on most observations.
However, this result does not exclude the presence of a small
fraction of lines of sight (of the order of a few percent) through
the most massive and/or the most metal-rich DLA systems and 
characterized by a large optical depth. Indeed, \citet{wil05} and
\citet{wil06} find a significant evidence of reddening in DLA
systems with CaII absorption lines at moderate redshift ($z_{abs}
\approx 1$; $\langle E(B-V) \rangle \gtrsim 0.1$). Similarly
\citet{yor06} measure $E(B-V)$ up to 0.085 for MgII selected DLA
systems at $z_{abs}\approx 1.0$. As the determination of the gas
density in DLA systems is dominated by these most massive absorbers,
the potential bias in this measure, correctly stressed by
\citet{fal93}, must not be dismissed by the recent evidence of a
very modest \emph{average} deviation from unity transmission.

In this paper we take advantage of the large sample of DLA systems
identified in Sloan quasars \citep{pro05} and we investigate the
relation between the observed and intrinsic density of neutral gas in
these systems. The Sloan sample that we consider has three main
advantages over the sample used by \citet{fal93}. (1) It is larger by
a factor 30. (2) The quasars have been selected within a luminosity
limit in the I band, significantly less sensitive to dust obscuration
than the B band used for the older sample. (3) The color selection
algorithm has a good sensitivity to red quasars \citep{ric02}. In fact
even if a quasar with a DLA absorption system is not dropped below the
I-band flux limit, its color can be changed so that it ends up lying
outside the color selection box used to identify quasar candidates for
follow-up spectroscopy. Thanks to the precision of the Sloan
photometry, that allows a clear separation of the stellar locus in the
color space, and to the use of an extended color selection box, the
loss of completeness due to this latter effect is only marginal (see
\citealt{ric02} for a detailed discussion of the Sloan color selection
algorithm and for completeness tests) and is therefore not considered
in this work.

This paper is organized as follows. In Sec.~\ref{sec:dust} we
characterize the obscuration bias in a magnitude limited survey, in
Sec.~\ref{sec:data} we describe the dataset that we are using. In
Sec.~\ref{sec:para_OM} we present our analysis for the parametric
estimation of the intrinsic comoving density of neutral gas, whose
uncertainties are quantified in Secs.~\ref{sec:boot}-\ref{sec:scatter}
by means of Monte Carlo simulations of synthetic observations. In
Sec.~\ref{sec:disc} we discuss the accuracy of a non-parametric
estimator for the neutral gas density. We summarize our findings in
Sec.~\ref{sec:conc}.

\section{Dust effects in magnitude limited surveys}\label{sec:dust}

In a magnitude limited survey, obscuration along the line of sight
leads to the potential loss of some lines of sight. Following
\citet{fal93}, to quantify the effect let us consider an intrinsic
quasar luminosity function $\Phi_i(L)$ in the interval
$[L_{min},+\infty]$, where $L_{min}$ is the luminosity above which the
sample is complete. The intrinsic number of objects above the
completeness threshold is:
\be
N_i = \int_{L_{min}}^{+\infty} \Phi_i(L) dL.
\ee
The presence of dust introduces an optical depth $\tau$ along the line
of sight, with a corresponding transmission coefficient $\exp{\{ -
\tau\}}$. Under this condition the number of objects with observed
luminosity above $L_{min}$ is:
\be
N_o = \int_{L_{min} e^{\tau }}^{+\infty} \Phi_i(L) dL.
\ee
%
The dust obscuration leads to a fraction of missing objects given by
$N_o/N_i$:
\be
\frac{N_o}{N_i} = \frac{ \int_{L_{min}e^{\tau }}^{+\infty} \Phi_i(L) dL}{ \int_{L_{min}}^{+\infty} \Phi_i(L) dL}
\ee
In general the precise value of $N_o/N_i$ for a given $\tau$ depends
on the sensitivity limit of the survey $L_{min}$ and on the form of
the luminosity function $\Phi(L)$. Under the assumption that the
luminosity function is a power law $\Phi(L) \propto L^{-\beta-1}$ (in
the luminosity range from $L_{min}$ to $+\infty$) the ratio $N_o/N_i$
can be easily computed and is independent of $L_{min}$:
\be \label{eq:bias}
\frac{N_o}{N_i} = \frac{L_{min}^{-\beta} e^{-\beta \tau }  }{L_{min}^{-\beta}} = e^{-\beta \tau }.
\ee
This was already noted by \citet{fal93}.

The luminosity function for quasars is modeled in terms of a gamma
function and/or of a double power law \citep[e.g.,
see][]{pei95}. However the SDSS spectroscopic quasar sample is
shallower, for $z_{qso} \gtrsim 2.2$, than the knee of the
distribution so that the luminosity function for the quasars that we
are considering in this paper can be effectively treated up to
$L_{min}$ as a single power law \citep[see][]{ric06}, simplifying
significantly our analysis by use of Eq.~(\ref{eq:bias}).

In the following analysis we assume a redshift dependent luminosity
function \citep{ric06}:
\be \Phi_i(L_i,z)=A(z) L_i^{-\beta(z)-1}, \ee
where $L_I$ is the luminosity in $I$ band\footnote{in this paper we
assume an average wavelength for the band of $\langle \lambda_I
\rangle =0.8 \mu m$.} (where the sensitivity limit for detection of
quasars in SDSS is given) and $\beta(z)$ is a slowly evolving function
of the redshift with $\beta$ varying from 2.2 to 1.1 in the redshift
range $[2;5]$ \citep{ric06}:
\be
\beta(z)= 2.1 - 0.275(z-2.45)\theta(z-2.45),
\ee
where $\theta(x)$ is the step function defined as $\theta(x)=1$ if $x
\geq 0$ and $\theta(x)=0$ otherwise. The SDSS DLA DR\_3 sample of
quasars considered in this paper has an average redshift of $2.97$
with a standard deviation of $0.65$. The average slope can be
approximated as:
\be
\langle \beta \rangle = 1.95 \approx 2.
\ee

In case of absorption due to a dusty DLA system at redshift $z_a$ with
column density $N_{HI}$, the optical depth at an observed wavelength
$\lambda_o$ can be written as:
\be \label{eq:tau}
\tau(\lambda_o,z_a,N_{HI},k) = k N_{HI} \xi(\lambda_o/(1+z_a)),
\ee
where $k$ is the dust-to-gas ratio and $\xi$ the relative extinction
curve normalized to the absorption cross section in B band
$\sigma(\lambda_B)$ \citep[e.g., see][]{pei92}:
\be
\xi(\lambda) = \sigma(\lambda)/\sigma(\lambda_B).
\ee
If $N_{HI}$ is expressed in units of $10^{21} cm^{-2}$, the absorption
cross section in B band is $kN_{HI}$ with $k=0.8$ for galactic
dust. The value of $k$ for DLA systems depends on their metallicity
$Z$ and on the fraction of metals in dust, i.e. on the dust-to-metals
ratio. DLA systems are generally characterized by a low metallicity
and by a moderate evolution of their properties with the redshift
\citep{wol05}. We approximate the observed average (HI-column density
weighted) redshift-metallicity relation (based on a number of
observations, e.g. \citealt{pro03}, \citealt{des04,des06},
\citealt{ake05}) with a linear function for $\log{(Z_o(z))}$. This
provides a good agreement with the data (e.g. see Fig. 13 in the
compilation by \citealt{kul05}) in the redshift range $2 \lesssim z
\lesssim 5$:
\be \label{eq:zav} Z_o(z)/Z_{\sun}=0.2 \cdot 10^{-0.2z}. \ee
The average metallicity in Eq.~(\ref{eq:zav}) translates, for a Milky
Way dust-to-metals ratio (i.e. $50\%$ of the metals in dust grains),
into a ``Milky Way'' average dust-to-gas ratio:
\be k_{MW}(z)=0.16 \cdot 10^{-0.2z}. \ee
To account for different fractions of metals in the form of dust
grains we introduce a correction factor $\alpha_{\kappa}$ for the
intrinsic dust-to-gas ratio $k_i(z)$:
\be \label{eq:kiz} k_i(z)=\alpha_{\kappa} k_{MW}(z).  \ee
DLA systems are considered to have a smaller fraction of metals in
dust than our galaxy (with roughly one quarter of the total metals
content in dust), so realistic values for $\alpha_{\kappa}$ are
expected to be around $\alpha_{\kappa} = 0.5$ (\citealt{pet97};
\citealt{vla02}; but see \citealt{pei99} where the Milky Way
dust-to-metals ratio is used). In the following sections we present
our analysis using a range of dust-to-metals ratio, highlighting the
dependence of the dust bias on this quantity. We assume a reference
value $\alpha_{\kappa}=0.5$ (25\% of metals in dust, the value derived
by \citealt{pet97}), but we include a grid of models with
$\alpha_{\kappa} \in [0;1]$ with $\Delta \alpha_{\kappa} = 0.125$,
discussing in depth also the case $ \alpha_{\kappa} = 1$ (50\% of
metals in dust).

Our analysis is carried out by assuming that all DLA at a given
redshift have a fixed metallicity, as well as a fixed dust-to-metals
ratio. A detailed modeling of the metallicity and dust-to-metals ratio
distributions for the purpose of determining $\Omega_{HI}^{(DLA)}$ is
extremely challenging, as only a small subsample of DLA system have
measured metallicities. Fortunately, the value of
$\Omega_{HI}^{(DLA)}$ is not expected to significantly depend on the
scatter in the dust-to-gas ratio $k_i$ of the absorbers (see
\citealt{fal93}, Appendix A). To ensure that this is the case, in
Sec.~\ref{sec:scatter} we validate our scatter-less approach by analyzing
synthetic observations with a variety of dust-to-gas distributions.
 
\section{The data: The DLA SDSS DR\_3 and the CORALS surveys}\label{sec:data}

The data used in this paper are primarily taken from the DLA survey by
\citet{pro05}. We consider all the quasars from their Table 1 and all
the DLA systems reported in their Table 3. The sample consists of 525
DLA system found in 4568 spectra of quasars with a minimum signal to
noise ratio of 4.

In addition we consider the DLA systems detections in the radio
selected CORALS survey \citep{ell01}, intrinsically free from dust
bias. This survey consists of 19 DLA systems detected in 66 spectra of
quasars (see their Table 3). The statistical sample for the survey
(see \citealt{ell01}), that we consider for the analysis, is
restricted to $17$ DLA in the redshift range $[1.8;3.5]$.

\section{Parametric Estimation of $\Omega_{HI}^{(DLA)}$ from SDSS DR3}\label{sec:para_OM}

Following the standard practice (e.g. \citealt{lan91}) we define the
number of observed DLA systems in the intervals
$[N_{HI},N_{HI}+dN_{HI}]$ and $[X,X+dX]$:
\be
f_{o}(N_{HI},X)dN_{HI}dX,
\ee
where $f_{o}$ is the observed frequency distribution and $dX$ the
absorption distance:
\be
dX \equiv \frac{H_0}{H(z)}(1+z)^2dz.
\ee
In this paper we adopt the standard $\Lambda CDM$ cosmology with
$H_0=70km/s/Mpc$, $\Omega_{\Lambda}=0.7$ and $\Omega_M=0.3$, so that:
\be dX = \frac{(1+z)^2}{\sqrt{0.7+0.3(1+z)^3}}dz.  \ee
The data from SDSS DLA DR\_3 span over a total integrated absorption
path-length $\Delta X = 7333.1$.

Considering the obscuration bias discussed in the previous section for
the Sloan survey (see also \citealt{fal93}) and assuming that the
distribution of absorbers along the line of sight is not
correlated\footnote{We are essentially neglecting clustering along the
line of sight.}, we can write the relation between the intrinsic and
the observed frequency distribution of DLA systems as:
\be \label{eq:fifo}
f_o(N_{HI},z)= f_i(N_{HI},X(z)) e^{-\langle \beta \rangle \alpha_{\kappa}k_{MW}(z) N_{HI} \xi(\lambda_I/(1+z))},
\ee
where we assume the average slope $\langle \beta \rangle=2$ for the
quasar luminosity function.  Following \citet{pro05} we parameterize
$f$ using a gamma function:
\be \label{eq:gamma}
f_i(N_{HI},X)= \eta_1 \left (\frac{N_{HI}}{N_{\gamma}} \right )^{m_1} \exp{\left (-\frac{N_{HI}}{N_{\gamma}} \right )},
\ee
and a power law:
\be \label{eq:power}
f_i(N_{HI},X)= \eta_2 \left ({N_{HI}} \right )^{m_2}.
\ee
The use of the gamma function for the intrinsic distribution $f_i$ has
the advantage that, given Eq.~(\ref{eq:fifo}), also the observed
distribution $f_o$ remains a gamma function. A intrinsic power law
is instead mapped into a observed gamma function by the effect of
dust.

If we assume that the intrinsic distribution of DLA systems follows a
Poisson statistics, the observed distribution will also follow a
Poisson statistics.
The likelihood to be maximized is therefore (see also Eq.~27 in
\citealt{fal93}):
\be \label{eq:like} \mathcal{L} =  \exp{-\left \{ \sum_s
\int_{{z_{inf}}_s}^{{z_{sup}}_s} dz \int_{N_{DLA}}^{+\infty} d
N_{HI} \cdot f_o({N_{HI}}_s,z_s) \frac{dX}{dz} \right \}} \prod_n
f_o({N_{HI}}_n,z_n), \ee
where the sum extends over all the sample of quasars that have been
searched for DLA systems in the redshift interval
$[{z_{inf}}_s;{z_{sup}}_s]$, where the signal to noise ratio is above
$4$, and the product is over all the DLA systems detected in those
quasars. In Eq.~(\ref{eq:like}) the observed distribution function
$f_o$ is expressed in terms of the intrinsic distribution $f_i$
through Eq.~(\ref{eq:fifo}). $N_{DLA}=2\cdot 10^{20} cm^{-2}$ is the
assumed lower limit for the column density of a DLA system.

Before presenting the results from the maximum likelihood analysis we
investigate the properties of $f_o$ with the aim to discuss upper
limits on the dust-to-gas ratio and on $\Omega_{HI}^{(DLA)}$.

\subsection{Basic considerations from the properties of $f_o$}

\begin{figure}
  \plotone{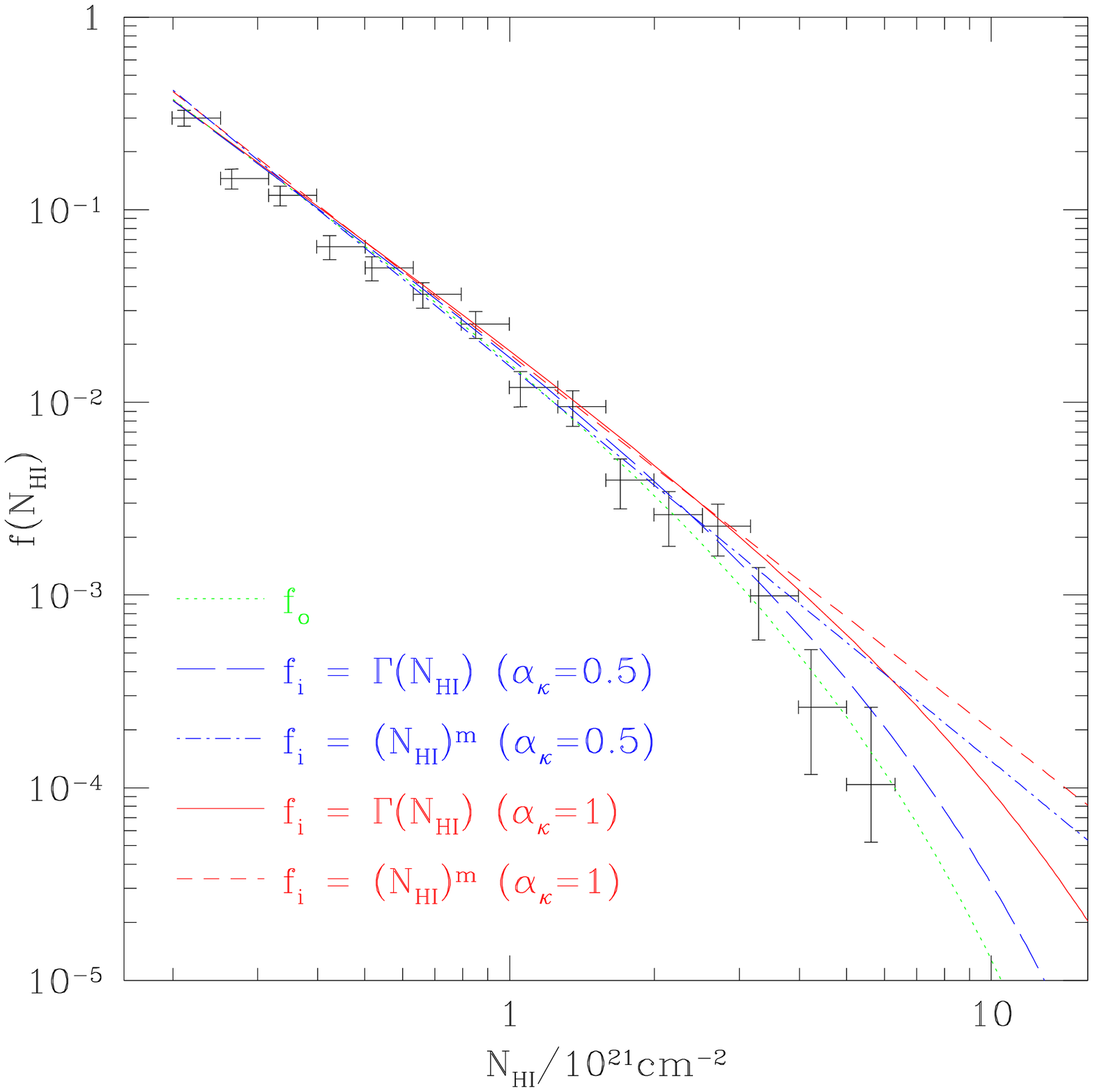} \caption{Observed column density distribution of
  DLA systems in the SDSS DLA DR\_3 sample compared to observed (solid)
  and intrinsic (dotted) best fitting gamma functions. The difference
  between $f_i$ and $f_o$ is due to dust obscuration, especially
  apparent for high column densities. The parameters value for the
  fitting functions are reported in
  Tables.~\ref{tab:likeGAMMA}-\ref{tab:likePOWER}.}\label{fig:coldensdis}
\end{figure}

Fig.~\ref{fig:coldensdis} shows the observed column density
distribution of DLA systems, averaged over redshift. The observed data
are well fitted by a gamma function (\citealt{pro05}, see also
Eq.~\ref{eq:gamma} and Tab.~\ref{tab:likeGAMMA}) with a knee at
$(N_{\gamma})_o=3 \cdot 10^{21} cm^{-2}$ and a slope $(m_1)_o=-1.8$.

The origin of the \emph{observed} knee at $(N_{\gamma})_o$ can in
principle be either due to an intrinsic decrease of the distribution
at high column densities, i.e. due to $N_{\gamma}$, or due to the
effect of obscuration bias, that introduces in the data an
exponential-like decrease. We can obtain a limit on the dust-to-gas
(expressed in terms of $\alpha_{\kappa}$) ratio assuming that the
observed knee is entirely due to dust bias. From Eq.~(\ref{eq:fifo})
we can write:
\be \langle \beta \rangle \alpha_{\kappa}k_{MW}(z) \xi(\lambda_I/(1+z))
\lesssim 1/(N_{\gamma})_o, \ee
so that, considering that the average redshift of DLA absorbers in our
sample is $\langle z \rangle = 3.1$ we estimate a limit on
$\alpha_{\kappa}$:
\be
\alpha_{\kappa} \lesssim  \frac{1}{(N_{\gamma})_o} \cdot \frac{1}{\langle \beta \rangle k_{MW}(\langle z \rangle) \xi(\lambda_I/(1+\langle z \rangle )) } \lesssim \frac{1}{3} \cdot \frac{1}{ 2 \cdot 0.038 \cdot 2.47} \lesssim 1.75.
\ee
%
As this value implies about $90\%$ of metals in dust, which appears
extremely improbable for DLA systems, the presence of an observed knee
in the column density distribution of the gas is likely an intrinsic
feature and not induced by the dust bias alone. This implies that the
expected difference between the intrinsic and the observed density of
neutral gas in DLA systems is limited for the most realistic fraction
of metals in dust ($25\%$ i.e. $\alpha_{\kappa}=0.5$). Only in the
case of a larger fraction of metals in dust (i.e. $\alpha_{\kappa}
\approx 1$) there can be the possibility of a significant dust bias,
as in that case an acceptable model of the intrinsic distribution can
still be obtained in terms of a power law.
In the next Section we quantify more precisely the dust bias by
studying the likelihood $\mathcal{L}$ for $f_i$ at increasing
$\Omega_{HI}^{(DLA)}$ for different values of $\alpha_{\kappa}$.

\subsection{Whole sample analysis}

\begin{figure}
  \plotone{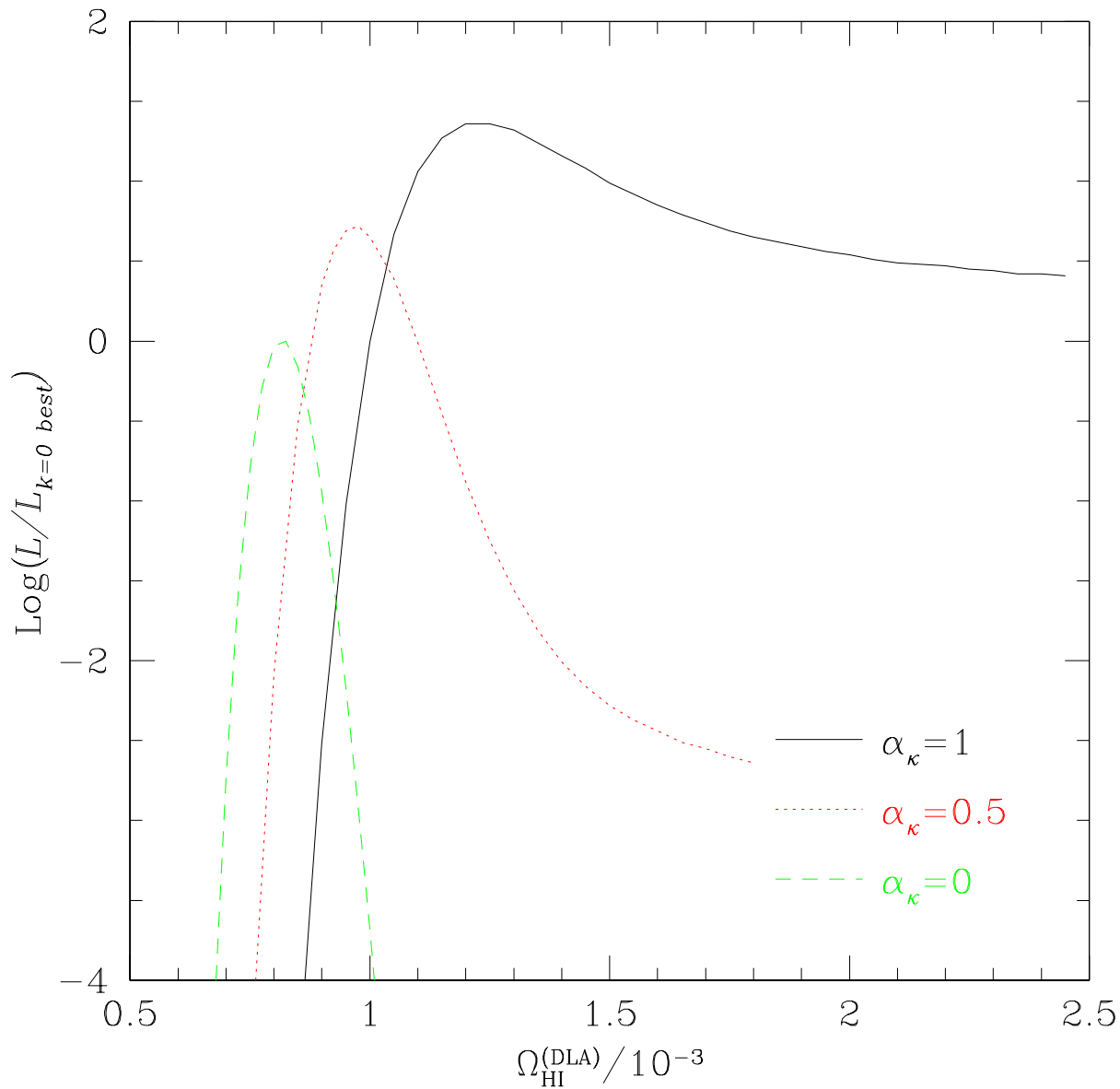} \caption{Maximum likelihood $\mathcal{L}$ for the
intrinsic comoving neutral gas density in DLA systems
$\Omega_{HI}^{(DLA)}$ for the SDSS DLA DR\_3 data (solid line with
$\alpha_{\kappa}=1$; red dotted with $\alpha_{\kappa}=0.5$ ). The
green dashed line is associated to the \emph{observed} gas density
($\alpha_{\kappa}=0$). The likelihood curves have been obtained by
maximizing $\mathcal{L}$ over $N_{\gamma}$ and $m_1$ at fixed
$\Omega_{HI}^{(DLA)}$ and $\alpha_{\kappa}$.}\label{fig:like}
\end{figure}

By aggregating all the data in SDSS DLA DR\_3, the results of our
maximum likelihood analysis are reported in Tables~\ref{tab:likeGAMMA}-\ref{tab:likePOWER} and in
Fig.~\ref{fig:like}.
The results for the fit with an intrinsic gamma function are reported
in terms of the comoving density of gas in DLA systems
(${\Omega_{HI}^{(DLA)}}$), that is the first moment of $f_i$:
\be \label{eq:om}
{\Omega_{HI}^{(DLA)}} dX = \frac{\mu m_H H_0}{c \rho_c}
\int_{N_{DLA}}^{+\infty} dN_{HI} N_{HI} f_i(N_{HI},z) dX,
\ee
where $m_H$ is the mass of the hydrogen atom, $\mu=1.3$ is a
correction factor for the composition of the gas, $c$ the speed of
light, $\rho_c$ the critical density and $N_{DLA}=2 \cdot 10^{20}
cm^{-2}$.

\begin{figure}
  \plottwo{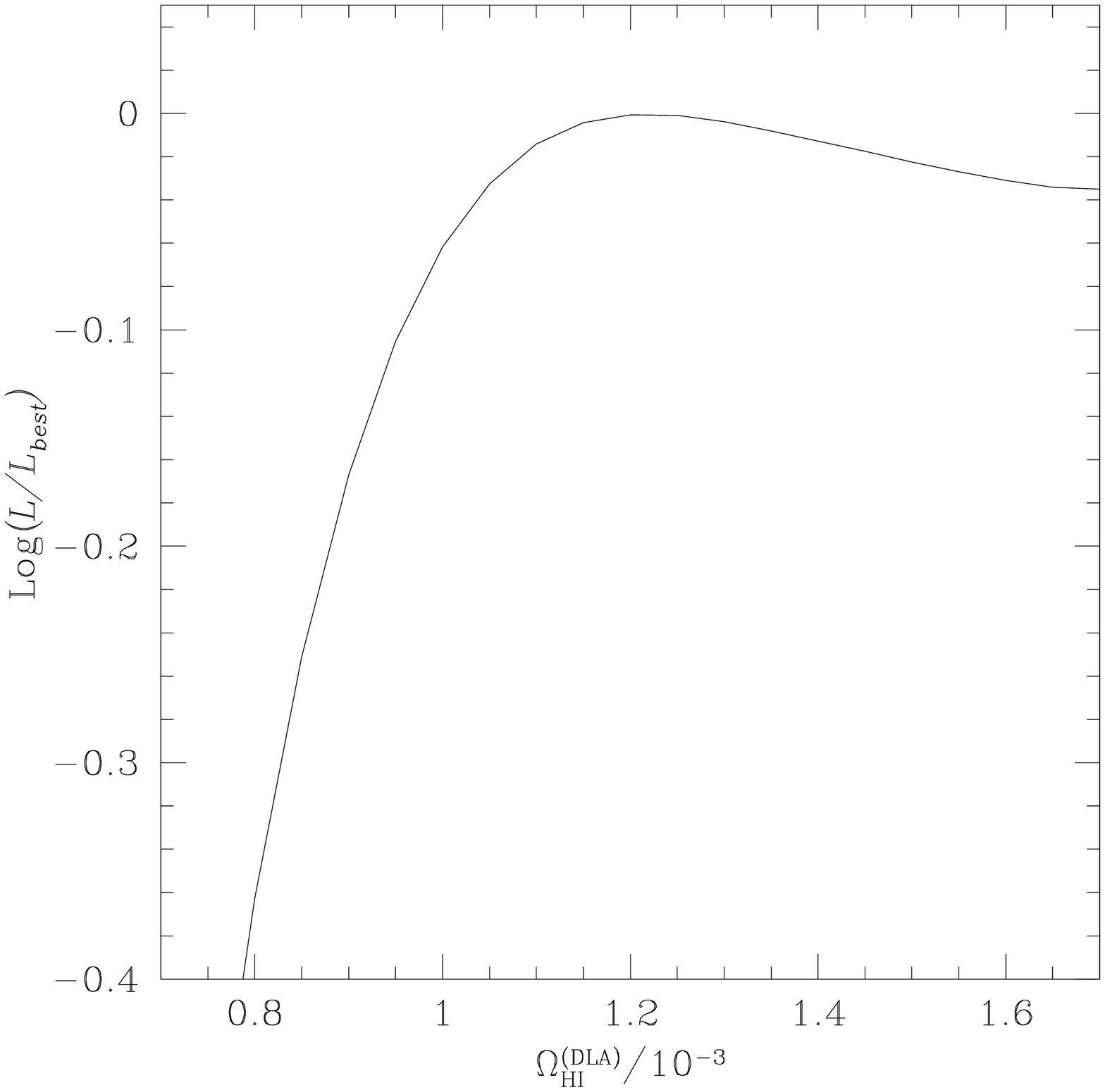}{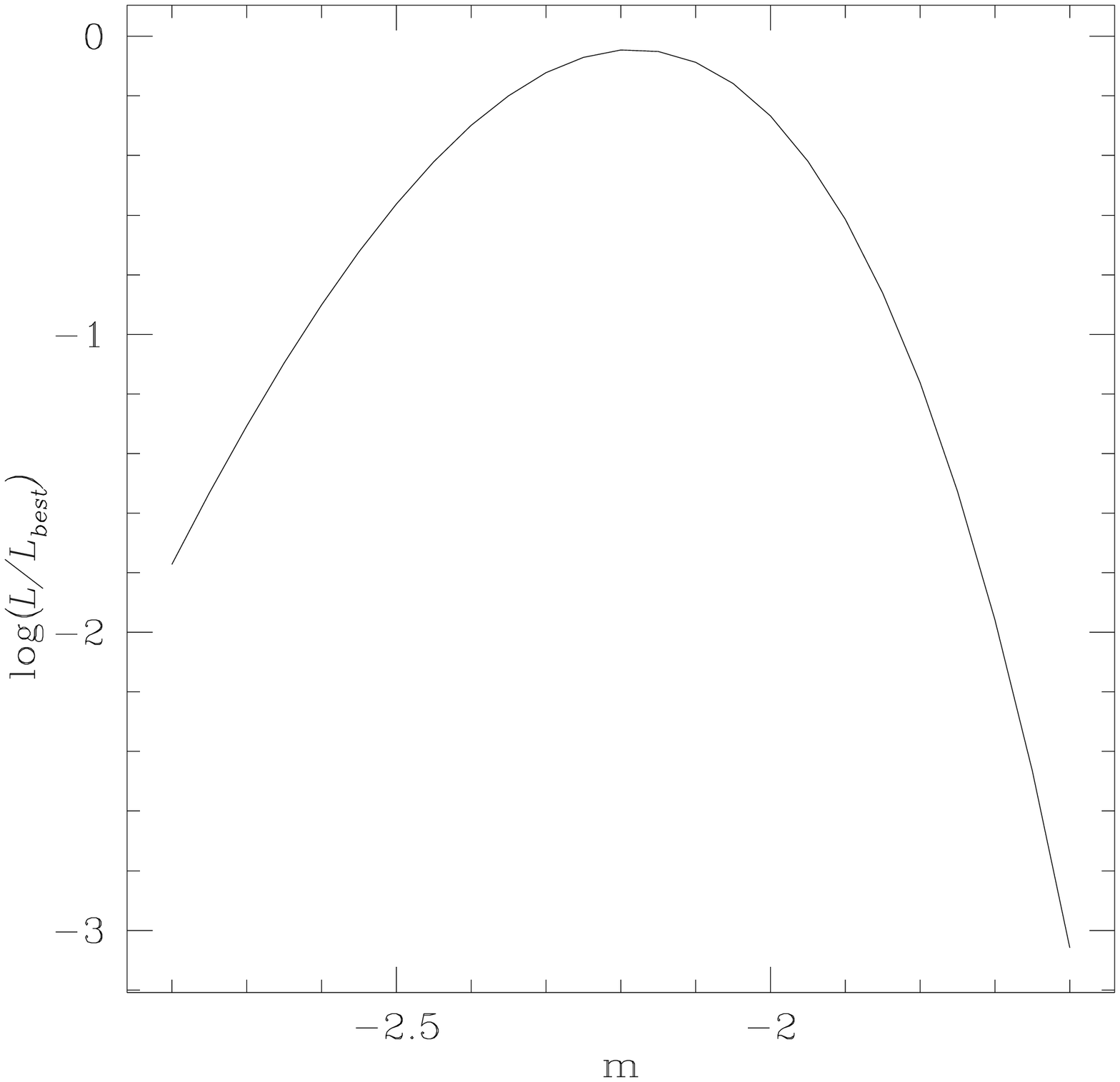} \caption{CORALS survey
analysis. Right: likelihood curve in function of $\Omega_{HI}^{(DLA)}$
for a gamma function modeling of the data. The curve shows only very
limited variations around the maximum value. This is because the
dataset is too small to allow to constraint the gamma function
parameters. Left: maximum likelihood $\mathcal{L}$ for the power law
slope $m_2$. The likelihood is in units of the maximum likelihood
value for the fit with a gamma function (see
Table~\ref{tab:likeCORAL}).}\label{fig:coral}
\end{figure}

We obtain the following results:
\begin{itemize}
 
\item {\bf $\alpha_{\kappa} = 0$}: If we neglect the effect of dust
absorption so that $f_o \equiv f_i$, we re-derive the parameters for
the gamma distribution reported in \citet{pro05} (see our
Tab.~\ref{tab:likeGAMMA}).  All our likelihood values for the Sloan
data are presented in log units normalized to the likelihood value for
the best fitting dust-free model. For $\alpha_{\kappa}=0$ the
likelihood is sharply peaked around the best value for
$\Omega_{HI}^{(DLA)}$. However, when we introduce the effect of the
dust ($\alpha_{\kappa}> 0$) there is an increasingly strong
degeneration toward high values for $\Omega_{HI}^{(DLA)}$ (see
Fig.~\ref{fig:like}).

\item {\bf $\alpha_{\kappa} = 0.5$}: For our standard scenario the
likelihood has a maximum at $\Omega_{HI}^{(DLA)}=0.97 \cdot 10^{-3}$,
to be compared with the value $\left(\Omega_{HI}^{(DLA)}\right
)_0=0.82 \cdot 10^{-3}$ given by the maximum likelihood analysis with
$\alpha_{\kappa}=0$ (no dust). In our modeling of the column density
distribution the metallicity is introduced has an external input,
based on independent observations. An additional support to motivate
the need of taking into account the obscuration bias comes from the
fact that the likelihood for $\alpha_{\kappa}=0.5$ is higher than for
$\alpha_{\kappa}=0$.

\item {\bf $\alpha_{\kappa} = 1$}: The bias induced by the dust
increases as larger $\alpha_{\kappa}$ are considered. As expected
qualitatively by the simple considerations on the shape of $f_o$
presented in the previous section, the maximum likelihood analysis for
our large dust-to-metal ratio scenario ($\alpha_{\kappa}=1$) is unable
to set a strong upper limit to $\Omega_{HI}^{(DLA)}$ as there is a
wide wing of high likelihood values toward large neutral gas
densities. At $\alpha_{\kappa}=1$ a power law $f_i$ with slope
$m_2=-1.95$ has a likelihood ratio over the best gamma solution of
$\mathcal{R} =
2\log({\mathcal{L}}_{\Gamma}/{\mathcal{L}}_{N^m})=2.1$. A likelihood
ratio test with this value implies that a power law solution can be
ruled out only at about 85\% of confidence level if we assume that
$\mathcal{R}$ is distributed as a $\chi^2$ with one degree of
freedom. However this is in general valid only when the model is
linear in the parameters and the errors follow a gaussian distribution
\citep{lup93}. As our model is strongly non linear, in the next
section we estimate the confidence level by means of Monte Carlo
simulations.

\item {\bf $\alpha_{\kappa} > 1$}: If we consider fits for $f_i$ at
$\alpha_{\kappa}> 0$ (see Table~\ref{tab:likePOWER}) we obtain an
increasingly better modeling in terms of power laws up to
$\alpha_{\kappa} \approx 2$. As expected from the preliminary
examination of $f_o$, eventually the likelihood decreases at higher
dust-to-gas ratio starting from $\alpha_{\kappa} \gtrsim 2$. We recall
that $\alpha_{\kappa}>2$ is formally unphysical, as it implies more
than 100\% of the metals in dust, however, due to the fact that we
have adopted a simple fit to the observed metallicity, a high
$\alpha_{\kappa}$ value also means that the intrinsic metallicity
content of DLA system is underestimated in our model. As it appears
unlikely that the metallicity measurements do significantly
underestimate the dust content of DLA systems, we expect that
realistic values of $\alpha_{\kappa} \leq 1$.

\item {\bf CORALS}: For comparison we have repeated the maximum
likelihood analysis on the radio selected CORALS quasars
\citep{ell01}. The results are reported in Table~\ref{tab:likeCORAL}
and in Fig.~\ref{fig:coral}.  For these data the measure is guaranteed
to be free from dust bias, but the sample size is insufficient to
accurately constraint the parameters of $f_i$ (the likelihood curve in
Fig.~\ref{fig:coral} is relatively flat around the maximum). The
maximum is at $\Omega_{HI}^{(DLA)}=1.2 \cdot 10^{-3}$ with extremely
large $1\sigma$ uncertainties (of the order of 50\%) due to the
flatness of the likelihood function. In addition, the maximum
likelihood for the gamma function model is only negligibly better than
the one for a power law ($\mathcal{R} \approx 0.1$).

\end{itemize}

\section{Errors estimate}\label{sec:boot}

\begin{figure}
  \plotone{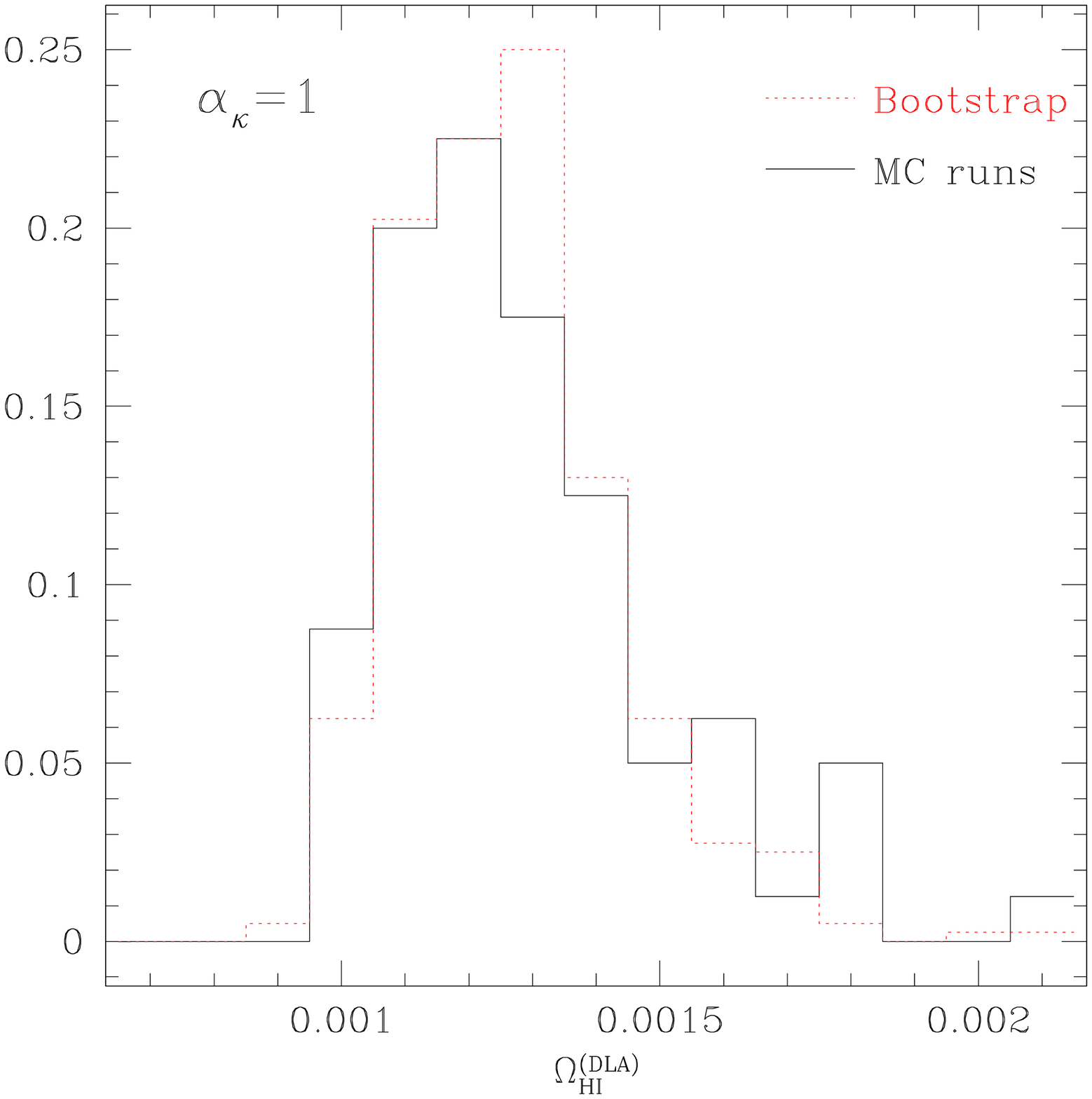} \caption{$\Omega_{HI}^{(DLA)}$ distribution
  for $\alpha_{\kappa}=1$ from the bootstrapping analysis of the
  data (red dotted line; 400 realizations) and from 80 Monte Carlo
  realizations (solid line) of synthetic observations starting from
  the best fitting model for $f_i$ (gamma function with $m_1=-1.8$,
  $N_{\gamma}=7.22 \cdot 10^{21} cm^{-2}$ and
  $\Omega_{HI}^{(DLA)}=1.25 \cdot 10^{-3}$).}\label{fig:omMC}
\end{figure}
\begin{figure}
  \plotone{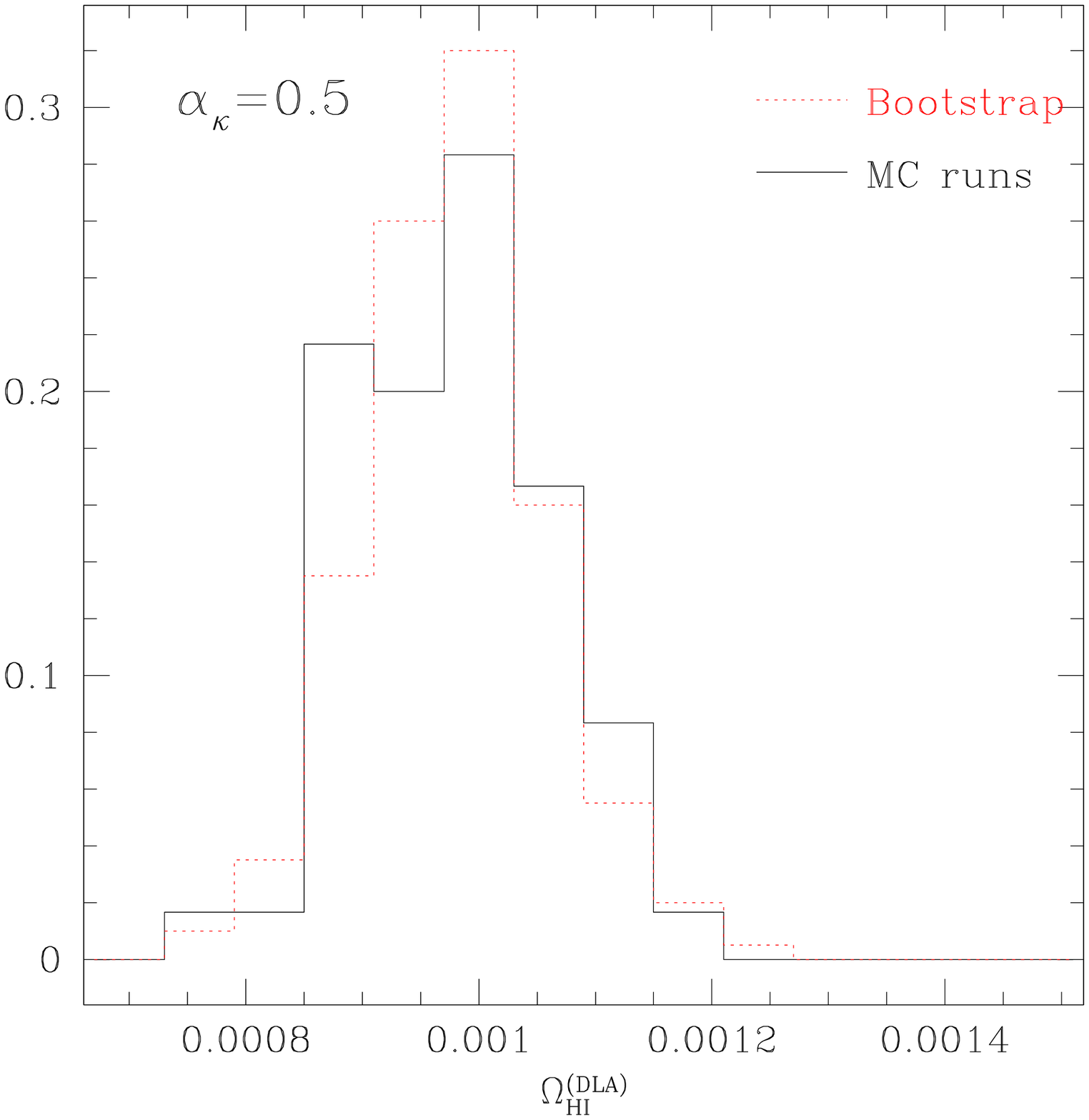} \caption{$\Omega_{HI}^{(DLA)}$ distribution
  for $\alpha_{\kappa}=0.5$ from the bootstrapping analysis of the
  data (red dotted line; 200 realizations) and from 60 Monte Carlo
  realizations (solid line) of synthetic observations starting from
  the best fitting model for $f_i$ (gamma function with $m_1=-1.79$,
  $N_{\gamma}=4.15 \cdot 10^{21} cm^{-2}$ and
  $\Omega_{HI}^{(DLA)}=0.97 \cdot 10^{-3}$).}\label{fig:omMCk05}
\end{figure}

To estimate the error on the determination of $\Omega_{HI}^{(DLA)}$ we
have performed a bootstrapping analysis of the data. We have randomly
extracted samples of lines of sight with uniform probability from the
SDSS DLA survey. The samples have the same number of lines as the original
data. For each of these simulated samples we have performed a maximum
likelihood analysis to determine $\Omega_{HI}^{(DLA)}$. In the case of
$\alpha_{\kappa}=1$, the results obtained with 400 random realizations
are presented in Fig.~\ref{fig:omMC} and show that
$\Omega_{HI}^{(DLA)}$ is indeed distributed around the best value
derived from the original data ($1.25 \cdot 10^{-3}$), which coincides
with the median of the distribution. The average value of the
distribution is $1.27 \cdot 10^{-3}$. At $68\%$ of confidence level
the measurements lie in the interval $[1.10 \cdot 10^{-3};1.41 \cdot
10^{-3}]$. We assume this interval as our fiducial error on
$\Omega_{HI}^{(DLA)}$. At $90\%$ of confidence level
$\Omega_{HI}^{(DLA)} \leq 1.48 \cdot 10^{-3}$. The bootstrapping
procedure has been repeated for all the scenarios with different
$\alpha_{\kappa}$, resorting to at least $100$ simulations for each
value of the dust-to-metals ratio considered. The results are reported
as standard $1\sigma$ error in Table.~\ref{tab:likeGAMMA}.  The case
$\alpha_{\kappa}=0.5$ is shown in Fig.~\ref{fig:omMCk05} (using 200
simulations in this case).


As a additional test to investigate the uncertainties in our analysis,
we have simulated a number of synthetic samples of data using the
Monte Carlo code developed in \citet{tre06}. These samples have then
been processed using the same method applied to the real observed
data, to test its efficiency in retrieving the input parameters and to
quantify the typical error.

Each random realization of a simulated observation is characterized by
the same observed path-length of the SDSS sample ($\Delta X = 7333$),
but consists of lines of sight with DLA distribution (and detection)
in the redshift range $[2.2;4]$\footnote{This has been done for
computational reason in order to speed up the evaluation of the double
integral in Eq.~\ref{eq:like}.} The optical depth along each intrinsic
line of sight is computed from the realized DLA distribution and then
a test on the obscuration probability (Eq.~\ref{eq:bias}) is performed
to accept or reject the intrinsic line of sight as a observed one. The
only source of uncertainty in these synthetic observations comes from
the discrete sampling of $f_i$ and from the obscuration probability
test. The model does not include observational errors.

We have simulated two models: (i) an intrinsic gamma function and (ii)
an intrinsic power law, with the parameters given by the maximum
likelihood analysis from the SDSS data (see Table~\ref{tab:likeGAMMA}
for the gamma function and Table~\ref{tab:likePOWER} for the power
law).

Fig.~\ref{fig:omMC} shows, for $\alpha_{\kappa}=1$, the distribution
of $\Omega_{HI}^{(DLA)}$ as measured from the simulated observations
(80 different realizations) of a intrinsic gamma function. It is
reassuring that the distribution basically agrees with the one
obtained using the bootstrap method. The mean value for
$\Omega_{HI}^{(DLA)}$ is $1.29 \cdot 10^{-3}$ with a standard
deviation of $0.23 \cdot 10^{-3}$. As we used an input value
$\Omega_{HI}^{(DLA)}=1.25\cdot 10^{-3}$ to generate the synthetic
observations, the fitting procedure is able to recover the intrinsic
density of neutral gas introducing only a negligible bias, which is
much smaller than the one sigma random uncertainty.

The analysis of the simulated observations generated from an intrinsic
power law distribution allows to quantify the probability that the
identification of a gamma function as best fitting model for the Sloan
data is only due to the combined effect of discrete sampling and of
the additional degree of freedom of the gamma function over a power
law. For our reference scenario ($\alpha_{\kappa}=0.5$) we have
generated 400 random realizations of observations starting from an
intrinsic power law with unconstrained neutral gas density
($\eta_2=1.54 \cdot 10^{-2}$, $m_2=-2.05$). The likelihood ratio
(gamma over power law) distribution shows only 2 realizations that
have $\mathcal{R}$ greater than the one measured from the SDSS DLA
data. This means that an intrinsic power law can be ruled out at
$99.5\%$ of confidence level. Repeating the experiment in the case of
$\alpha_{\kappa}=1$, $\eta_2=1.77 \cdot 10^{-2}$ and $m_2=-1.95$ we
find that about $8~\%$ of the realizations have a value of
$\mathcal{R}$ greater than the one measured from the SDSS DLA data.
This means that the confidence level at which a power law distribution
(with unconstrained upper cut-off) can be ruled out is in this case $92\%$.

\section{Synthetic observations with dust-to-gas ratio scatter}\label{sec:scatter}

In the maximum likelihood applied in this work we assume that all the
absorbers have a fixed dust-to-metals ratio $\alpha_{\kappa}$ and a
metallicity dependent only on the redshift. In reality a significant
scatter around the mean values is expected for both these quantities
and has indeed been observed \citep[e.g. see][]{kul05}. \citet{fal93}
have already noted that the scatter in the dust-to-gas ratio is not
introducing significant bias under the assumption that there is no
correlation between $k_{i}$ and the column density $N_{HI}$.

To confirm the results of \citet{fal93} and to extend the analysis of
the bias to the case where a correlation between $k_{i}$ and $N_{HI}$
is present we have simulated a number of synthetic samples of data
adopting an approach similar to the one described in
Sec.~\ref{sec:boot}. In this case we compute the optical depth of each
absorber by drawing dust-to-gas ratio $k_i$ from a given distribution,
rather than adopting a fixed value depending only on $z$.

When there is no correlation between $k_i$ and $N_{HI}$ we use (based on
Eq.~\ref{eq:kiz}):
\be
k_i(z)= \xi \cdot \alpha_{\kappa} \cdot k_{MW}(z),
\ee
where $\xi$ is a random variable with a given probability distribution
$p(\xi)$ such that the expected value of $\xi$ under $p$ is
$E_{p}[\xi]=1$. We have considered the following form for $p(\xi)$:
\begin{itemize}
\item A uniform distribution around the average value: $p(\xi)=1$ if
$\xi \in [0.5;1.5]$ and $p(\xi)=0$ otherwise.
\item A lognormal distribution with variance parameter $\sigma_*$, as
studied by \citet{fal93}: $p(\xi)=\exp{\{-\log^2{(\xi
\exp{\{\sigma_*^2/2\}})}}/(2\sigma^2_*)\}/\sqrt{2 \pi \sigma^2_*
\xi^2}$. We adopt $\sigma_*=0.5,1,1.5$ inspired by the dispersion in
the dust-to-gas ratio observed for galaxies in the local universe
\citep{pei92}.
\end{itemize}
In addition we have generated synthetic observations with a
correlation between $k_i$ and $z_i$ adopting the following
prescription: 
\be
k_i(z)= \xi \cdot N_{HI}^{\gamma} \frac{\int_{N_{DLA}}^{+\infty} dN_{HI}\cdot N_{HI} f_i(N_{HI})}{\int_{N_{DLA}}^{+\infty} dN_{HI}\cdot N_{HI}^{1+\gamma} f_i(N_{HI})} \cdot \alpha_{\kappa} \cdot k_{MW}(z),
\ee
where $\gamma = -0.8,-0.4,0.4,-0.8$ (i.e we explore both a increasing
and decreasing trend of $k_i$ vs. $N_{HI}$) and $\xi$ is a random
variable with the same properties bescribed above. The normalization
adopted is such that the HI column density averaged value of $k_i$ is
$\alpha_{\kappa} \cdot k_{MW}(z)$.

\begin{figure}
  \plotone{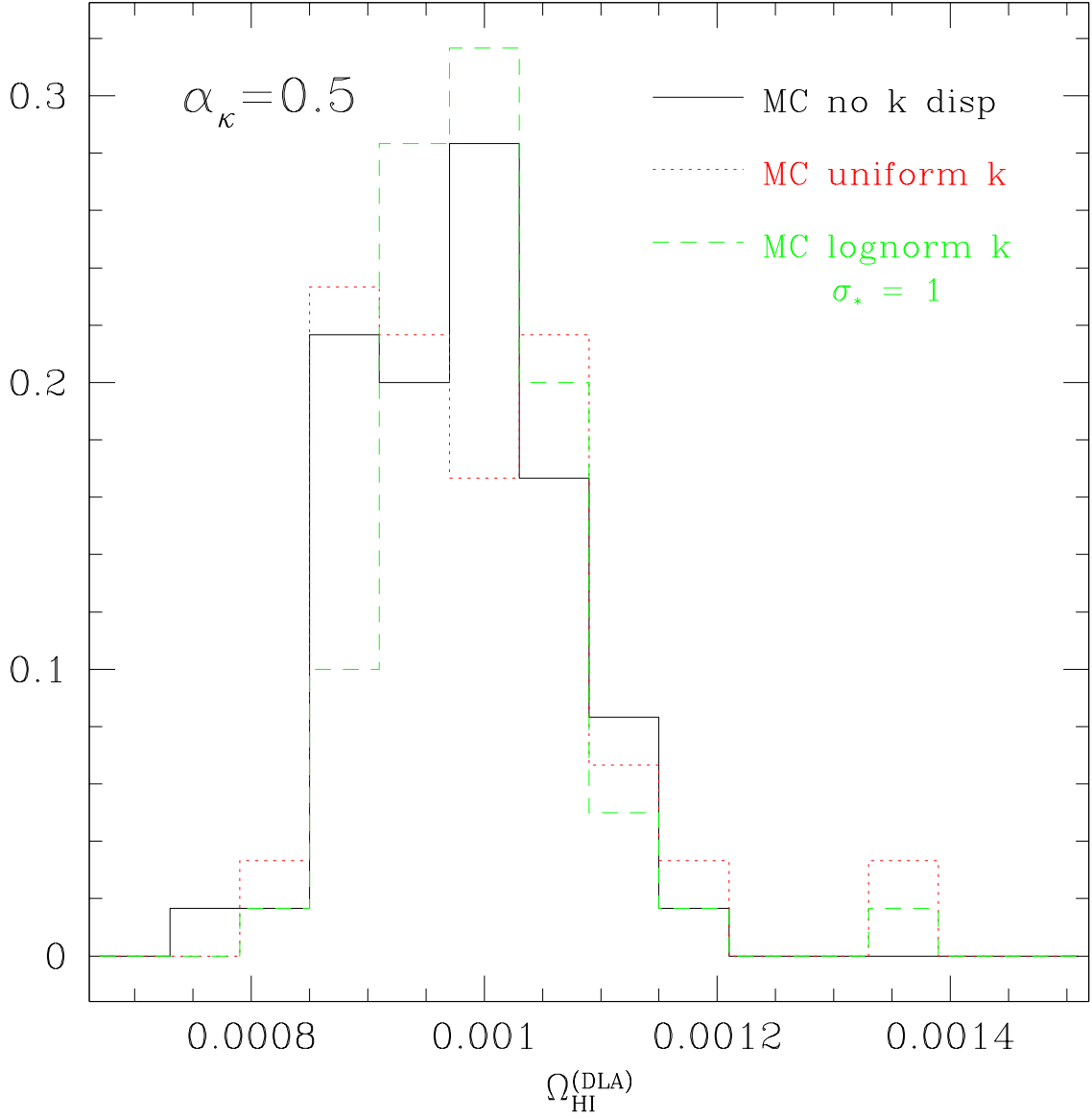} \caption{$\Omega_{HI}^{(DLA)}$ distribution from
  Monte Carlo realizations with a dust-to-gas ratio scatter of
  synthetic observations starting from the best fitting model for
  $f_i$ (gamma function with $m_1=-1.79$, $N_{\gamma}=4.15 \cdot
  10^{21} cm^{-2}$ and $\Omega_{HI}^{(DLA)}=0.97 \cdot 10^{-3}$). The
  histograms have been constructed using 60 random realizations
  each. The solid black line is for a model with no scatter, the red
  dotted line for a model with a uniform dispersion in $k_i$ and the
  green dashed line for a model with a lognormal
  distribution (with $\sigma_*=1$).}\label{fig:scatter}
\end{figure}

\begin{figure}
  \plotone{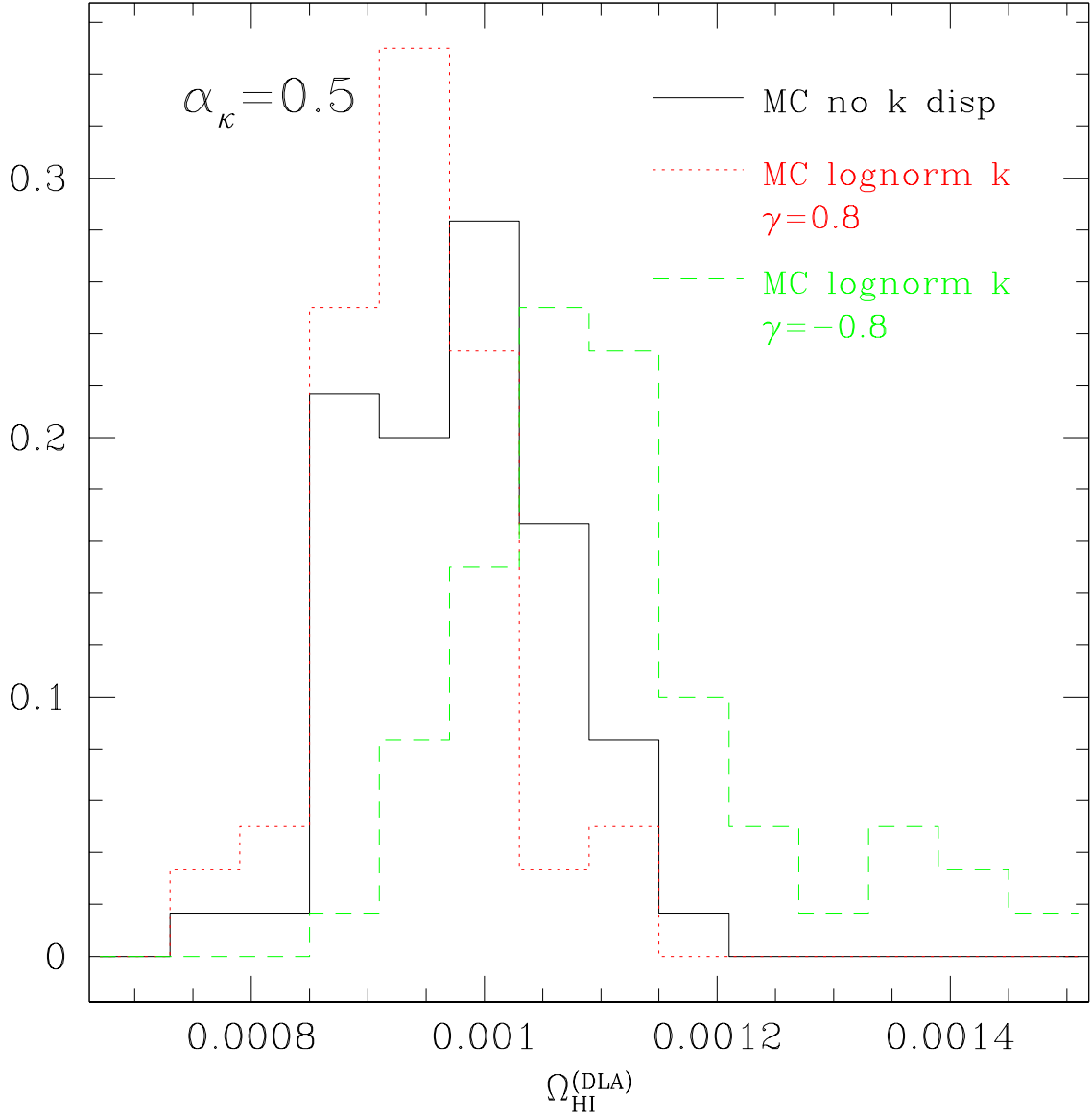} \caption{Like fig.~\ref{fig:scatter}, but for
  synthetic observations with a lognormal $k_i$ distribution
  ($\sigma_*=1$) and a large correlation ($\gamma= \pm 0.8$) between
  $k_i$ and $N_{HI}$ (green dashed and red dotted lines)
  vs. realizations with no scatter (black solid
  line).}\label{fig:scatterCOR}
\end{figure}

For each combination of the parameters considered (see
Table~\ref{tab:scatter}), we have generated 60 synthetic SDSS-like
samples of data (as in Sec.~\ref{sec:boot}) adopting a gamma function
model for $f(N_{HI})$ ($N_{\gamma}=4.15 \cdot 10^{21} cm^{-2}$;
$m_1=-1.8$) and $\alpha_{\kappa}=0.5$. These synthetic samples have
all $\Omega_{HI}^{(DLA)}=0.97 \cdot 10^{-3}$. By applying our maximum
likelihood we obtain the following results (summarized in
Table~\ref{tab:scatter}):
\begin{itemize}
\item Uncorrelated scatter in $k_i$. The maximum likelihood analysis
is able to correctly recover the value of $\Omega_{HI}^{(DLA)}$ used
to generate the observations (see Fig.~\ref{fig:scatter} and
Table~\ref{tab:scatter}). There is only a small bias (at the percent
level) toward obtaining output values marginally larger than the
input, even when the dust obscuration is such that the observed column
density averaged dust-to-gas ratio is smaller by $15\%$ with respect
to the intrinsic value, as happens in our lognormal model with
$\sigma_*=1$. This demonstrates that our analysis in terms of the
average dust-to-gas ratio is robust with respect to uncorrelated
scatter in $k_i$. Interestingly we also note that the difference
between the observed and the intrinsic dust-to-gas ratio for the
lognormal model with $\sigma_*=1$ is in agreement with the difference
between the metallicity in optically and radio selected samples of DLA
systems (see \citealt{ake05})\footnote{Note however that the
difference in the metallicities measured by \citet{ake05} is not
particularly robust from a statistical point of view, as it is
within the $1\sigma$ error bar.}.
\item Correlated scatter in $k_i$. In this scenario the analysis in
terms of an average dust-to-gas ratio systematically mis-estimates the
dust content at both ends of the column density distributions of the
absorbers. In particular, if $k_i$ increases with $N_i$, then
absorbers with the largest column densities will be more dust rich
than assumed by our analysis. This means that these absorbers will
more likely be dropped by the sample than is assumed in the modeling,
so that $\Omega_{HI}^{(DLA)}$ is underestimated. When $k_i$ decreases
with $N_i$, the opposite happens: the densest absorbers are less dusty
than assumed and $\Omega_{HI}^{(DLA)}$ is overestimated. Due to the
skewness of the lognormal distribution the bias introduced by the
correlation is expected to be larger when $k_i$ is a decreasing
function of $N_{HI}$. This is indeed what we observe in the analysis
of our synthetic observations (see Fig.~\ref{fig:scatterCOR} and
Table~\ref{tab:scatter}). The bias introduced is, however, relatively
modest (at most $15\%$ in $\Omega_{HI}^{(DLA)}$), even assuming a
rather strong dependence of $k_i$ on $N_{HI}$ (i.e. $k_i \propto
N_{HI}^{\pm 0.8}$). The bias could be avoided by taking into account
the precise form of the correlation into the modeling of the the
dust-to-gas ratio. Unfortunately the $k_i$ vs. $N_{HI}$ correlation
appears hard to quantify with the present data. Only a small subset of
the known DLA systems have a measured metallicity and highly dusty and
dense absorbers are preferentially missed from the sample. In addition
there are some evidences that the fraction of metals depleted in dust
depends on the column density of the absorbers
\citep{wel97}. Therefore plots of the measured gas phase metallicity
versus the hydrogen column density, such as those presented in
\citet{boi98} and \citet{sav00} cannot be easily interpreted to
extract the $k_i(N_{HI})$ relation.
\end{itemize}

\section{Non-parametric Estimation of $\Omega_{HI}^{(DLA)}$ from SDSS DR3}\label{sec:disc}

The comoving density of DLA systems can be also estimated using a
non-parametric approach by taking the discrete limit of
Eq.~(\ref{eq:om}) combined with Eq.~(\ref{eq:fifo}):
\be \label{eq:omegaDIS}
{\Omega_{HI}^{(DLA)}} = \frac{\mu m_H H_0}{c \rho_c} \frac{\sum_s {N_{HI}}_se^{\beta_s \tau_s}}{(\Delta X)_i}.
\ee
The sum is done over all the DLA systems in the sample, while the
intrinsic total path-length $(\Delta X)_i$ is given by summing over all
the individual path-lengths for the quasars in the survey, with a weight
depending on the optical depth $\tau_n$ along the line of sight to the
$n-th$ quasar and on the redshift-dependent slope of the quasar
luminosity function $\beta(z)$:
\be
(\Delta X)_i = \sum_n (dX_o)_n e^{\beta_n \tau_n}
\ee
The optical depth $\tau_n$ is given by:
\be
\tau_n = \alpha_{\kappa} \sum_p k_{MW}(z_p) {N_{HI}}_p \xi(\lambda_I/(1+z_p)),
\ee
where the index $p$ is running over all the DLA systems along the
$n-th$ line of sight.

The random uncertainty on the measure has been estimated by using a
`Jack-knife' analysis \citep{lup93}: we derive ${\Omega_{HI}^{(DLA)}}$
for $N=22$ subsamples of quasars by ignoring $N_{tot}/N$ quasars each
time. The uncertainty on $\Omega_{HI}^{(DLA)}$ is then given by:
\be \sigma^2 = \frac{N-1}{N} \sum_s (\Omega_s - \langle \Omega_s
\rangle)^2,  \ee
where $\Omega_s$ stands for ${\Omega_{HI}^{(DLA)}}_s$.

\begin{figure}
  \plotone{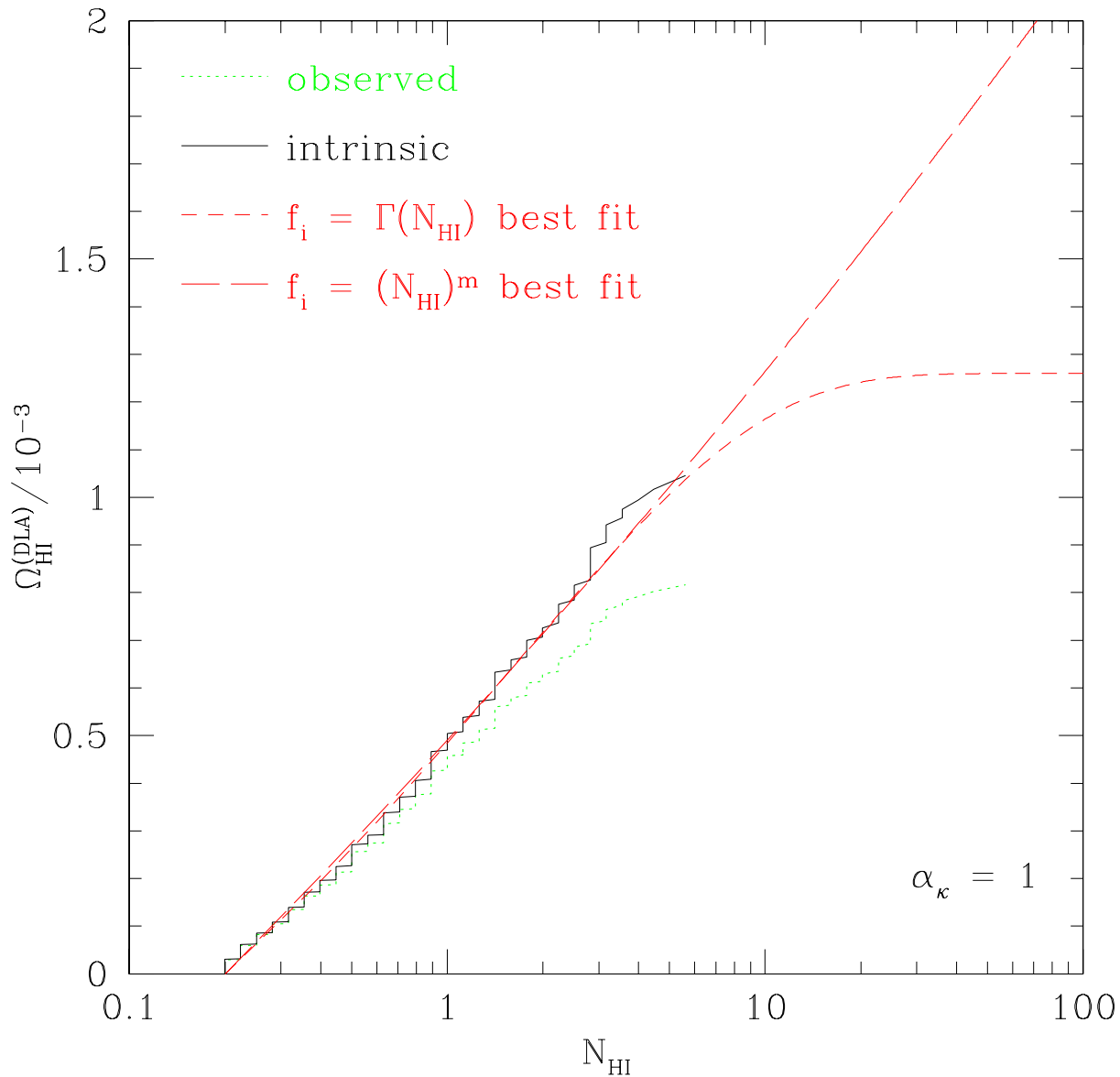} \caption{Cumulative neutral gas density in DLA
systems versus column density for the SDSS DLA DR\_3 (\emph{observed}
dotted, \emph{intrinsic} with $\alpha_{\kappa}=1$ solid). Superimposed
to the intrinsic curve we show the profile from the best fitting gamma
function $f_i$ (short dashed) and from a power law $f_i$ (long dashed)
with slope $-1.95$: the discrete evaluation for the \emph{intrinsic}
distribution does not extend to sufficiently high column densities to
allow convergence.  }\label{fig:cumul}
\end{figure}

We obtain the following results: 

\begin{itemize}

\item {\bf $\alpha_{\kappa}=1$}: Eq.~\ref{eq:omegaDIS}, applied to the
whole sample with $\alpha_{\kappa}=1$, gives $\Omega_{HI}^{(DLA)}=1.04
\cdot 10^{-3}$ with an error of $8 \cdot 10^{-5}$ (see
Table~\ref{tab:discrete}).  The obscuration bias based on this estimate
is of about $30~\%$.  This represents a lower limit on the bias, as
from Fig.~\ref{fig:cumul} it is evident that with $\alpha_{\kappa}=1$
the dust bias is so severe that the discrete evaluation of
${\Omega_{HI}^{(DLA)}}$ has not converged. On the top of the random
errors, estimated with the 'Jack-knife' method, this non parametric
estimation suffers from systematic effects of the order of at least
$10-20\%$.

\item {\bf $\alpha_{\kappa}=0.5$}: The values of $\Omega_{HI}^{(DLA)}$
obtained in our reference model with $\alpha_{\kappa}=0.5$ for
different redshift intervals are reported in
Tab~\ref{tab:discrete}). The systematic corrections due to the dust
bias are relatively modest (of the order of 10\%) but still represent
a contribution to the total uncertainty that cannot be neglected. Like
in the case discussed above a systematic effect is also present. Here
the magnitude of the effect is smaller, of the order of 5\%, as can be
evaluated from the difference in $\Omega_{HI}^{(DLA)}$ between this
method and the maximum likelihood analysis of the previous section.

\item {\bf CORALS}: We also apply the same analysis on the DLA
detections in the CORALS survey \citep{ell01}, obtaining
${\Omega_{HI}^{(DLA)}}= (1.2\pm 0.5) \cdot 10^{-3}$ ($1\sigma$ error
bars; see Table~\ref{tab:discrete}). Note that our analysis leads to a
different value for $\Omega_{HI}^{(DLA)}$ because in \citet{ell01} the
analysis was performed with a different cosmology ($\Omega_M=1$,
$\Omega_{\Lambda}=0$). The main limit of the survey is its modest
path-length extension ($\Delta X = 200.8$, to be compared with the
Sloan pathlength $\Delta X =7333.1$), which is translated into a large
random error on $\Omega_{HI}$. In addition potential systematics
errors due to incompleteness may bias the CORALS measure much like
dust introduces a bias in optically selected surveys. In fact the
likelihood modeling of the data is inconclusive, allowing a variety of
different forms for $f_i$ (see Fig.~\ref{fig:coral}).

\end{itemize}

\section{Conclusions}\label{sec:conc}

In this paper we have improved the analysis of the column density
distribution of DLA systems in the SDSS DR\_3 DLA survey\citep{pro05}
to take into account the bias due to dust obscuration along the line
of sight. A first modeling of the bias was constructed by
\citet{ost84} and improved by \citet{fal93}. These earlier
estimates expected a severe effect of obscuration, with the observed
gas density of DLA systems being up to several times smaller than the
intrinsic one. The best estimate for $\Omega_{HI}^{(DLA)}$ given by
\citet{fal93} is $4.9 \cdot 10^{-3}$ (obtained using a cosmology with
$\Omega_M=1$ and $H_0=70 km/s/Mpc$) and their upper limit
$\Omega_{HI}^{(DLA)} \leq 3 \cdot 10^{-2}$, obtained from a standard
big bang nucleosynthesis abundance model. More recent works
\citep{mur04,ell05} tend to dismiss the issue of obscuration bias
based on the absence of systematic reddening in the spectra of quasars
with DLA absorption features.

Here we show that the effect of obscuration, while not being as severe
as predicted by \citet{fal93}, does indeed play an important effect on
the precise measurement of $\Omega_{HI}^{(DLA)}$. In the era of
precision cosmology, where the \emph{observed} density
${\Omega_{HI}^{(DLA)}}$ is constrained with errors below $10~\%$, the
systematic effects are not to be underestimated. With the typical
amount of metals present in DLA systems the \emph{observed} density
${\Omega_{HI}^{(DLA)}}$ derived from shallow magnitude limited surveys
of quasars underestimates the \emph{intrinsic} density
${\Omega_{HI}^{(DLA)}}$ by about $15~\%$ assuming dust-poor DLA
systems (i.e. systems with a fraction of metals in dust of 25\%). Our
best estimation for $z \in [2.2;5.5]$ gives an intrinsic neutral gas
density $\Omega_{HI}^{(DLA)}=0.97^{+0.08}_{-0.06} \cdot 10^{-3}$
($1\sigma$ error bars) to be compared with the observed gas density
$\Omega_{HI}^{(DLA)}=0.817^{+0.050}_{-0.052} \cdot 10^{-3}$ derived by
\citet{pro05}. If we leave the dust to metal ratio parameter
$\alpha_{\kappa}$ free to vary over the relevant range (from 0 to 1),
we find $\Omega_{HI}^{(DLA)}=0.97^{+0.08+0.28}_{-0.06-0.15} \cdot
10^{-3}$, where the first set of error bars gives the $1\sigma$ random
errors and the second set gives the modeling uncertainty dependent on
the fraction of metals in dust.  The obscuration bias therefore
represents the main source of uncertainties for the determination of
the intrinsic neutral gas content in DLA systems.

Our analysis has been carried out assuming that all the DLA absorbers
at a given redshift have the same dust-to-gas ratio. By means of monte
carlo simulations of synthetic observations we shoow that this
assumption does not introduce a significant bias as long as the
dust-to-gas ratio is not correlated with the hydrogen column
density. A correlation of the form $k_i \propto N_{HI}^{\gamma}$
introduces a systematic error of the order $+14\%$ in
$\Omega_{HI}^{(DLA)}$ for $\gamma = - 0.8$ and of $-3 \%$ for $\gamma
= +0.8$ (the bias is reduced to $+8\%$ and $0\%$ for $\gamma = \mp
0.4$ respectively).

Caution is also needed when taking the result from the maximum
likelihood at face value. In our reference scenario we find that a
power law function for column density distribution of DLA systems can
be ruled out at a confidence level no greater than 99.5\%. If the dust
content in DLA systems is higher, the dust bias becomes more
significant. For $50\%$ of the metals in dust grains the present data
do not allow to put an upper limit to the neutral gas content with
confidence level greater than $90\%$. We show in fact that there is a
probability of about 10\% that the SDSS DLA DR\_3 data are consistent
with an intrinsic power law distribution in column density adopting
$\alpha_{\kappa}=1$. The slope of this power law is $m_2=-1.95$ and an
upper cut-off cannot be derived from the data. In optically selected
surveys, absorbers with high column densities of neutral gas and with
high metallicity are missed, letting systematic uncertainties go
easily out of control.

Radio selected quasar samples would represent an elegant, bias free
solution to the measurements of $\Omega_{HI}^{(DLA)}$.
Unfortunately, with the present extension, these surveys constraint
$\Omega_{HI}^{(DLA)}$ with a random uncertainty of the order of
$40~\%$ (plus potential systematic effects due to the limited number
of observed DLA systems). To reduce the errors within the $10~\%$
level without the need to assume a modeling for the dust bias it is
therefore necessary to increase the path-length of radio surveys by
a factor $10$ at least. The UCSD radio survey \citep{jor06} has
recently made promising progresses in this direction.

\acknowledgements

It is a pleasure to thank Michael S. Fall for stimulating discussions
and Jason X. Prochaska for useful suggestions and comments on a draft
of the paper. We are grateful to the referee for constructive
suggestions. This work was supported in part by NASA JWST IDS grant
NAG5-12458.


\clearpage

\clearpage

\clearpage
\clearpage
\clearpage
\clearpage

\clearpage

\clearpage

\clearpage

\begin{table}
\begin{center}
\caption{Maximum likelihood parameters for gamma function $f_i$\label{tab:likeGAMMA}}
\begin{tabular}{l|ccccc}
\tableline\tableline
$\alpha_{\kappa}$ & $\eta_1$ & $m_1$ & $N_{\gamma}$ &$log(\mathcal{L})$ & $\Omega_{HI}^{(DLA)}$ \\
0.000 & $3.03~10^{-3}$ & $-1.80$ & $3.00$ & $0.00$ & $0.82^{+0.05}_{-0.05}~10^{-3}$ \\
0.125 & $2.61~10^{-3}$ & $-1.80$ & $3.23$ & $0.16$ & $0.85^{+0.05}_{-0.05}~10^{-3}$ \\
0.250 & $2.49~10^{-3}$ & $-1.78$ & $3.40$ & $0.32$ & $0.89^{+0.05}_{-0.05}~10^{-3}$ \\
0.375 & $2.21~10^{-3}$ & $-1.78$ & $3.61$ & $0.46$ & $0.91^{+0.07}_{-0.05}~10^{-3}$ \\
0.500 & $1.70~10^{-3}$ & $-1.79$ & $4.15$ & $0.66$ & $0.97^{+0.08}_{-0.06}~10^{-3}$ \\
0.625 & $1.47~10^{-3}$ & $-1.79$ & $4.51$ & $0.83$ & $1.01^{+0.10}_{-0.08}~10^{-3}$ \\
0.750 & $1.10~10^{-3}$ & $-1.79$ & $5.27$ & $1.00$ & $1.08^{+0.12}_{-0.10}~10^{-3}$ \\
0.875 & $1.02~10^{-3}$ & $-1.78$ & $5.60$ & $1.18$ & $1.13^{+0.13}_{-0.12}~10^{-3}$ \\
1.000 & $6.24~10^{-4}$ & $-1.80$ & $7.22$ & $1.33$ & $1.25^{+0.16}_{-0.15}~10^{-3}$ \\

 \tableline
\end{tabular}
\tablecomments{Summary table with the maximum likelihood parameters
for the intrinsic distribution of DLA systems fitted with a gamma
function and using different values for $\alpha_{\kappa}$.  The
parameters $\eta_1$ and $N_{\gamma}$ are given adopting units of
$10^{21} cm^{-2}$ for $N_{HI}$. The likelihood $\mathcal{L}$ is
normalized to the value of the best fitting model (gamma function)
with no dust.}
\end{center}
\end{table}


\clearpage
\begin{table}
\begin{center}
\caption{Maximum likelihood parameters for a power law function
$f_i$\label{tab:likePOWER}}
\begin{tabular}{l|ccc}
\tableline\tableline
$\alpha_{\kappa}$ & $\eta_2$ & $m_2$ & $log(\mathcal{L})$ \\
0.000 & $1.27~10^{-2}$ & $-2.19$ & $-9.90$ \\
0.125 & $1.36~10^{-2}$ & $-2.14$ & $-7.07$ \\
0.250 & $1.43~10^{-2}$ & $-2.10$ & $-5.26$  \\
0.375 & $1.49~10^{-2}$ & $-2.07$ & $-3.84$   \\
0.500 & $1.54~10^{-2}$& $-2.05$ & $-2.70$ \\
0.625 & $1.61~10^{-2}$ & $-2.01$ & $-1.73$ \\
0.750 & $1.67~10^{-2}$ & $-1.99$ & $-1.95$ \\
0.875 & $1.74~10^{-2}$ & $-1.96$ & $-0.24$ \\
1.000 & $1.77~10^{-2}$ & $-1.95$ & $0.29$ \\
1.250 & $1.91~10^{-2}$ & $-1.90$ & $1.19$ \\
1.500 & $2.05~10^{-2}$ & $-1.85$ & $1.82$ \\
2.000 & $2.35~10^{-2}$ & $-1.75$ & $2.39$ \\
4.000 & $3.75~10^{-2}$ & $-1.45$ & $-1.12$ \\
 \tableline
\end{tabular}
\tablecomments{Summary table with the maximum likelihood parameters
for the intrinsic distribution of DLA systems fitted with a power
law function and using different values for $\alpha_{\kappa}$.  The
parameters $\eta_1$ and $N_{\gamma}$ are given adopting units of
$10^{21} cm^{-2}$ for $N_{HI}$. The likelihood $\mathcal{L}$ is
normalized to the value of the best fitting model (gamma function)
with no dust.}
\end{center}
\end{table}


\clearpage

\begin{table}
\begin{center}
\caption{Maximum likelihood parameters for the CORALS survey \label{tab:likeCORAL}}
\begin{tabular}{ccccc}
\tableline\tableline
$\eta_1$ & $m_1$ & $N_{\gamma}$ &$log(\mathcal{L})$ & $\Omega_{HI}^{(DLA)}$ \\
$6.33~10^{-6}$ & $-2.13$ & $39.09$ & $0$ & $1.2~10^{-3}$\\
\tableline
$\eta_2$ & $m_2$ & & $log(\mathcal{L})$ \\
$1.87~10^{-2}$ & $-1.80$ &  &$-1.15$ \\
$1.47~10^{-2}$ & $-2.20$ &  &$-0.05$ \\
$1.11~10^{-2}$ & $-2.50$ &  &$-0.55$ \\
 \tableline
\end{tabular}
\tablecomments{Summary table with the maximum likelihood parameters
for the CORALS survey modeling. The likelihood value is normalized
to the value given by the best fitting CORALS model. Other units are
like in Table~\ref{tab:likeGAMMA}}
\end{center}
\end{table}


\clearpage

\begin{table}
\begin{center}
\caption{Synthetic observations with scatter in the dust-to-gas ratio
\label{tab:scatter}}
\begin{tabular}{lccc}
\tableline\tableline
Model & $10^3 \langle \Omega_{HI}^{(DLA)} \rangle$ & $10^3 \sigma(\Omega_{HI})$ & $10^3 \Delta \Omega_{HI}$\\
\tableline
Uniform 50\%-150\% & 1.00 & 0.11 & 0.03 \\
LogNor $\sigma_*=0.5$ & 0.97 & 0.08 & 0.00 \\
LogNor $\sigma_*=1$ & 0.99 & 0.09 & 0.02 \\
LogNor $\sigma_*=1.5$ & 1.03 & 0.10 & 0.05 \\
\tableline
LogNor $\sigma_*=1$; $\gamma=+0.4$ & 0.97 & 0.10 & 0.00 \\
LogNor $\sigma_*=1$; $\gamma=-0.4$ & 1.05 & 0.10 & 0.08 \\
LogNor $\sigma_*=1$; $\gamma=+0.8$ & 0.94 & 0.07 & -0.03 \\
LogNor $\sigma_*=1$; $\gamma=-0.8$ & 1.11 & 0.13 & 0.14 \\
 \tableline
\end{tabular}
\tablecomments{Summary of the synthetic observations with a scatter in
the dust-to-gas ratio $k_i$. The underlying column density
distribution is the best fitting SDSS gamma function when
$\alpha_{\kappa}=0.5$ ($N_{\gamma}=4.15 \cdot 10^{21} cm^{-2}$,
$m_1=-1.79$, $\Omega_{HI}^{(DLA)}=0.97 \cdot 10^{-3}$). The first
series of models employs a uncorrelated dispersion in $k_i$ (uniform
and lognormal, see Sec.~\ref{sec:scatter}), the second series also
assumes a correlation between $k_i$ and $N_{HI}$. For each model we
have analyzed 60 random realizations to recover $\Omega_{HI}^{(DLA)}$
reporting the mean value obtained from the maximum likelihood analysis
with average $k_i(z)$ (second column), the standard deviation (third
column) and the bias, i.e. the average value minus the nominal model
value (fourth column).}
\end{center}
\end{table}


\clearpage

\begin{table}
\caption{Discrete evaluation of $\Omega_{HI}^{(DLA)}$ \label{tab:discrete}}
\begin{tabular}{l}
\tableline \tableline
SDSS:~~~~~ \\
\tableline \tableline
\end{tabular}
\begin{tabular}{lllllll}
\tableline
$z_{min}$ & $z_{max}$ & $dX_o$ & $N_{DLA}$ & $10^3 (\Omega_{HI}^{(DLA)})_{\alpha_{\kappa}=1}$  & $10^3 (\Omega_{HI}^{(DLA)})_{\alpha_{\kappa}=0.5}$ & $10^3 (\Omega_{HI}^{(DLA)})_{\alpha_{\kappa}=0}$ \\
\tableline
  2.2 &   5.5 &  7333.1 &  525 & $1.04 \pm 0.08$ &  $0.91 \pm 0.07$ & $0.82 \pm 0.06$ \\ 
  2.2 &   2.5 &  1589.2 &  83 &  $0.53 \pm 0.07$ &  $0.48 \pm 0.06$ &$0.45 \pm 0.06$  \\ 
  2.5 &   3.0 &  2697.6 &  189 & $1.04 \pm 0.14$ &  $0.90 \pm 0.11$ & $0.79 \pm 0.09$ \\ 
  3.0 &   3.5 &  1836.3 &  152 & $1.27 \pm 0.18$ &  $1.13 \pm 0.15$ &$1.01 \pm 0.13$ \\ 
  3.5 &   4.0 &   877.3 &  69 &  $1.47 \pm 0.30$ &  $1.27 \pm 0.24$ &$1.11 \pm 0.19$ \\ 
  4.0 &   5.5 &   332.6 &  33 &  $1.13 \pm 0.31$ &  $1.06 \pm 0.27$ & $1.00 \pm 0.24$ \\ 
\tableline
\end{tabular}
\begin{tabular}{l}
 \\
\tableline \tableline
CORALS: \\
\tableline \tableline
\end{tabular}
\begin{tabular}{lllll}
\tableline
$z_{min}$ & $z_{max}$ & $dX_o$ & $N_{DLA}$ & $10^3 \Omega_{HI}^{(DLA)}$ \\
\tableline
  1.8 &   3.5 & 200.8 & 17 & $1.20 \pm 0.50$  \\ 
\tableline

\end{tabular}
\tablecomments{$\Omega_{HI}^{(DLA)}$ evaluated in the discrete limit
for different redshift intervals from the SDSS and CORALS data. The
first two columns give the redshift interval considered, the third the
corresponding pathlength probed by the data ($dX_o$). $N_{DLA}$ is the
number of systems identified in the interval, $\Omega_{HI}^{(DLA)}$
the intrinsic density of neutral gas (evaluated using
$\alpha_{\kappa}=1$, $\alpha_{\kappa}=0.5$ and $\alpha_{\kappa}=0$). The uncertainty is
evaluated with a 'Jack knife' analysis. }

\end{table}

\end{document}